\pdfoutput=1
\documentclass[fleqn,usenatbib,usedcolumn]{mnras}

\usepackage{graphicx,amssymb,hyperref}
\usepackage{multirow}
\usepackage{color}

\usepackage[fleqn]{amsmath}
\usepackage{abbrev}
\usepackage{siunitx}
\usepackage{comment} 
\usepackage{gensymb}
\usepackage{booktabs,tabularx}
\usepackage{flushend}
\usepackage{xspace}
\usepackage{physics}
\usepackage{xcolor}

\usepackage{tikz}
\usetikzlibrary{math}
\usetikzlibrary{patterns}

\usepackage{nicefrac}
\usepackage{units}
\usepackage{verbatim}
\usepackage[shortlabels]{enumitem}

\setlist[itemize]{leftmargin=5pt}

\def\mnras{MNRAS}
\def\apj{ApJ}

\def\simlt{\lower.5ex\hbox{$\; \buildrel < \over \sim \;$}}
\def\simgt{\lower.5ex\hbox{$\; \buildrel > \over \sim \;$}}
\def\etal{{\it et al.}}

\def\kpc{\mathrm{\, kpc}}

\def\mpc{\mathrm{\, Mpc}}
\def\gpc{\mathrm{\, Gpc}}

\def\msun{\mathrm{\, M_\odot}}

\def\zs{z^\mathrm{S}}
\def\zl{z^\mathrm{L}}

\newcommand{\eagle}{\textsc{eagle}\xspace}

\newcommand{\bahamas}{\textsc{bahamas}\xspace}

\def\gs{\mathrel{\raise1.16pt\hbox{$>$}\kern-7.0pt \lower3.06pt\hbox{{$\scriptstyle \sim$}}}}         
\def\ls{\mathrel{\raise1.16pt\hbox{$<$}\kern-7.0pt \lower3.06pt\hbox{{$\scriptstyle \sim$}}}}   

\newcommand{\vect}[1]{\boldsymbol{#1}}

\newcommand{\orcid}[1]{\href{https://orcid.org/#1}{\includegraphics[scale=0.08]{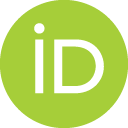}}}

\defcitealias{Hilbert2008}{H08}

\newcommand{\be}{\begin{equation}}
\newcommand{\ee}{\end{equation}}
\newcommand{\ba}{\begin{eqnarray}}
\newcommand{\ea}{\end{eqnarray}}

\DeclareGraphicsExtensions{.pdf,.png}

\title[What does strong lensing?]{What does strong gravitational lensing? The mass and redshift distribution of high-magnification lenses}

\author[A.\ Robertson \etal]
{\parbox{\textwidth}{Andrew Robertson$^1$\thanks{e-mail: {\tt andrew.robertson@durham.ac.uk}}\orcid{0000-0002-0086-0524}, 
, Graham P. Smith$^2$\orcid{0000-0003-4494-8277}
, Richard Massey$^1$\orcid{0000-0002-6085-3780}
, Vincent Eke$^1$\orcid{0000-0001-5416-8675}
, Mathilde Jauzac$^{1,3,4}$\orcid{0000-0003-1974-8732}
, Matteo Bianconi$^2$\orcid{0000-0002-0427-5373}  
and Dan Ryczanowski$^2$\orcid{0000-0002-4429-3429}
}
\vspace{0.3cm}
\\$^1$Institute for Computational Cosmology, Durham University, South Road, Durham DH1 3LE, UK
\\$^2$School of Physics and Astronomy, University of Birmingham, Birmingham, B15 2TT, UK
\\$^3$Centre for Extragalactic Astronomy, Department of Physics, Durham University, Durham DH1 3LE, UK
\\$^4$Astrophysics and Cosmology Research Unit, School of Mathematical Sciences, University of KwaZulu-Natal, Durban 4041, South Africa
}

\begin{document}

\maketitle

\label{firstpage}

\begin{abstract}

Many distant objects can only be detected, or become more scientifically valuable, if they have been highly magnified by strong gravitational lensing. We use \eagle and \bahamas, two recent cosmological hydrodynamical simulations, to predict the probability distribution for both the lens mass and lens redshift when point sources are highly magnified by gravitational lensing. For sources at a redshift of two, we find the distribution of lens redshifts to be broad, peaking at $z \approx 0.6$. The contribution of different lens masses is also fairly broad, with most high-magnification lensing due to lenses with halo masses between $10^{12}$ and $10^{14} \msun$. Lower mass haloes are inefficient lenses, while more massive haloes are rare. We find that a simple model in which all haloes have singular isothermal sphere density profiles can approximately reproduce the simulation predictions, although such a model over-predicts the importance of haloes with mass $<10^{12} \msun$ for lensing. We also calculate the probability that point sources at different redshifts are strongly lensed. At low redshift, high magnifications are extremely unlikely. Each $z=0.5$ source produces, on average, $\num{5e-7}$ images with magnification greater than ten; for $z =2$ this increases to about $\num{2e-5}$. Our results imply that searches for strongly lensed optical transients, including the optical counterparts to strongly lensed gravitational waves, can be optimized by monitoring massive galaxies, groups and clusters rather than concentrating on an individual population of lenses.

\end{abstract}
\begin{keywords}
gravitational lensing: strong, cosmology: theory, gravitational waves
\end{keywords}

\section{Introduction}

Gravitational lensing and gravitational waves are two phenomena predicted by Einstein's General Theory of Relativity (GR), both of which have now been observed. Evidence for strong gravitational lensing of electromagnetic radiation was first reported by \citet{1979Natur.279..381W}, who discovered a multiply imaged quasar. Since then strong gravitational lensing has become a key tool in astrophysics, allowing the mass distribution in galaxies and galaxy clusters to be mapped \citep[e.g.][]{2008ApJ...682..964B, 2010MNRAS.404..325R}, and for distant objects to be studied that would otherwise be too faint and/or small \citep[see][for a spectacular example]{2018NatAs...2..334K}. Gravitational waves were first detected only recently \citep{2016PhRvL.116f1102A} but also hold promise as a new tool for investigating our Universe \citep{2019Natur.568..469M}.

The prediction of GR is that gravitational waves have their trajectories bent by gravity in the same way as light, that is to say that gravitational waves can also be gravitationally lensed. Ignoring the lensing of gravitational waves could lead to incorrect conclusions about the population of merging compact objects. To leading order, the frequency evolution, $\nu(t)$, of a gravitational wave signal is determined by the chirp mass, $\mathcal{M} = (m_1 m_2)^{3/5} / (m_1 +m_2)^{1/5}$, where $m_1$ and $m_2$ are the masses of the two merging objects \citep{2017AnP...52900209A}. The intrinsic amplitude of the emitted signal can be predicted from the chirp mass, so the measured amplitude of the signal can be used to determine the luminosity distance, $d_\mathrm{L}$, to the coalescing objects.

For objects at redshift $z$, the received frequency will be lower than the intrinsic frequency by a factor of $1+z$. With an assumed cosmology and ignoring peculiar velocities, there is a one to one relationship between $d_\mathrm{L}$ and $z$. This means that the chirp mass and luminosity distance (and therefore also redshift) can be inferred from the amplitude and frequency evolution of a gravitational wave signal. Gravitational lensing introduces a new quantity, the magnification $\mu$, which alters the amplitude of the detected gravitational wave strain by a factor of $\sqrt{\mu}$ compared with the no lensing ($\mu = 1$) case. A gravitational wave signal which -- ignoring lensing -- would correspond to a chirp mass $\mathcal{M}_{\mu=1}$ at a redshift $z_{\mu=1}$, could actually be at a true redshift $z$, with true chirp mass $\mathcal{M}$ where $\mathcal{M} = \mathcal{M}_{\mu=1} (1+z_{\mu=1}) / (1+z)$. The magnification in this case would be $\mu = \left[d_\mathrm{L}(z) / d_\mathrm{L}(z_{\mu=1}) \right]^2$. As such, low mass, distant, and highly magnified gravitational wave sources, will masquerade as more massive and closer gravitational wave sources if gravitational magnification is not considered in interpreting the gravitational wave signal.

At LIGO/Virgo's current sensitivity, the detection rate of strongly lensed GWs is expected to be small \citep{2018MNRAS.476.2220L, 2018MNRAS.475.3823S}.  This can be understood in terms of the lens magnification that is required to render a lensed GW detectable.  At present, unlensed GWs can only be detected out to a luminosity distance of $\sim1 \gpc$ \citep{2018LRR....21....3A}, corresponding to $z \simeq 0.2$.  Given that the typical redshift of a strong lensing galaxy or cluster is $z \simeq 0.2-0.5$, it is inevitable that the typical redshift of a lensed GW is $z\gs1$, which corresponds to a luminosity distance of $\gtrsim 7 \gpc$.  Given the dependence of lens magnification on luminosity distance discussed above, these distances imply $\mu \gtrsim 50$, and thus a low probability of detection thus far, bearing in mind that $P(>\mu) \propto \mu^{-2}$.  Despite some early proof of concept follow-up observations \citep{2019MNRAS.485.5180S} and discussion of LIGO's early detections \citep{Broadhurst2018, 2019arXiv190103190B}, there is scant evidence that any of LIGO/Virgo's detections to date are strongly lensed \citep{2019ApJ...874L...2H, 2019arXiv191003601S}.

Calculating the expected number of detectable strongly lensed GWs, and the probability that a given GW detection is strongly lensed requires a good understanding of the population of gravitational lenses.  Knowledge of this population can then also shape the observing strategy for electromagnetic follow-up of candidate lensed GWs.  There has been disagreement in the literature as to what population of lenses is most important for the high magnifications required to reinterpret the LIGO/Virgo events as being less massive but at high redshift. For example, \citet{2018MNRAS.475.3823S} used the results of ray-tracing through a large $N$-body simulation \citep[][hereafter, \citetalias{Hilbert2008}]{Hilbert2008} to conclude that galaxy clusters were the most important population of lenses. Meanwhile, \citet{2018arXiv180707062H} forward modelled a population of lensed gravitational wave signals, assuming that lensed images dominantly arise due to galaxy lenses, modelling their population of lenses on SDSS early-type galaxies \citep{2007ApJ...658..884C}. Claims that some of the gravitational wave signals already detected are likely strongly lensed \citep{Broadhurst2018, 2019arXiv190103190B} have also assumed that galaxies dominate the optical depth for strong lensing, which they justify from the population of lenses responsible for lensed high redshift star forming galaxies detected by the Herschel satellite \citep{2010Sci...330..800N, 2013ApJ...762...59W, 2013ApJ...779...25B}. However, other lensed objects paint a different picture, with the first detected strongly lensed supernova \citep{2015Sci...347.1123K} and highly magnified individual star \citep{2018NatAs...2..334K}, as well as the most magnified lensed quasars \citep{2005ApJ...629L..73S, 2013MNRAS.429..482O, 2017ApJ...835....5S}, being lensed by galaxy clusters. Inferring the importance of different lenses for strong lensing using the distribution of lenses responsible for observed highly magnified objects is complicated by selection effects, which motivates answering this question from a theoretical perspective.

In this paper we seek to answer the question of which objects are responsible for producing strong gravitational lensing. Given the small size of the GW emission region in compact binary coalescence, and the high-magnifications required to reinterpret gravitational wave events as having come from objects significantly less massive than originally inferred, we will concentrate on high-magnifications ($|\mu| > 10$) for point sources. We do this by combining lensing calculations performed on two state of the art hydrodynamical simulations, and show that the results can be approximately reproduced by a model in which the total mass distribution of each gravitationally collapsed object is modelled as a singular isothermal sphere. 

This paper is organised as follows. In Section~\ref{sect:SIS} we describe a simple model for the strong lensing optical depth, based upon combining the halo mass function with a simple density profile for each halo. Then in Section~\ref{sect:lensing_in_hydro} we describe our simulations and how we calculated the cross-sections for strong gravitational lensing from individual simulated haloes. In Section~\ref{sect:results} we present our results, combining the strong lensing properties of all our simulated haloes to find the contribution of different lens masses and lens redshifts to the probability for strong lensing. We then present our conclusions in Section~\ref{sect:conc}. We assume a \citet{2014A&A...571A...1P} cosmology throughout this paper,\footnote{With $\Omega_\mathrm{m}=0.307$, $\Omega_\mathrm{b}=0.04825$, $\Omega_\mathrm{\Lambda}=0.693$, $\sigma_8 = 0.8288$, $n_\mathrm{s} = 0.9611$ and $h = 0.6777$.}  unless stated otherwise.

\section{A simple model for the strong lensing optical depth}
\label{sect:SIS}

Answering the question of `\emph{what does strong lensing?}' requires two ingredients. First, the \emph{lensing cross-section} of individual lenses as a function of their mass and redshift, and second, the \emph{halo mass function} -- the number density of haloes of different masses at different cosmic times. In this section we discuss both of these in the context of a simple model where all lenses are assumed to be spherically symmetric \emph{singular isothermal spheres} (SISs), which have 3D density profiles with $\rho \propto r^{-2}$. The advantage of using this density profile is that it provides a simple analytic method to relate the Einstein radius of a lens to the mass of its dark matter halo, across a broad range of halo mass, as detailed in Sections~\ref{sect:sigma_lens} and \ref{sect:hmf}.  The SIS model has been shown to be a good description of strong lensing galaxies \citep{2007ApJ...667..176G}. Whilst this model is a poor description of individual strong-lensing clusters due to the complexity of such systems \citep[e.g.][]{2010MNRAS.404..325R}, the slope of the SIS profile is representative of cluster density profile slopes on the relevant scales and is therefore appropriate for population studies such as this. The goal of this section is to introduce notation and gain intuition that will help to understand the results from hydrodynamical simulations in Section~\ref{sect:lensing_in_hydro}.

\subsection{Lensing cross-sections of individual lenses}
\label{sect:sigma_lens}

For a mass distribution at redshift $\zl$, we can define the gravitational lensing cross-section for some criterion as the solid angle that satisfies that criterion. An example criterion could be that the magnification, $\mu$, for a source at redshift $\zs$ is greater than some magnification threshold, $\mu_0$. Gravitational lensing maps an infinitesimal solid angle in the source plane, into a solid angle in the image plane that differs by a factor of the magnification.\footnote{The image plane solid angle does not need to be larger, as $|\mu|$ can be less than unity, but throughout this paper we are concerned with high magnifications.} As such, we can define cross-sections both in the image plane and the source plane, and in general these can be different. To introduce our notation we take a concrete example of a cross-section for magnification greater than 10 for a redshift 2 source. A particular lens would then have an image plane cross-section, $\sigma_\mathrm{lens}^\mathrm{I}(|\mu|>10, \zs=2)$, and source plane cross-section, $\sigma_\mathrm{lens}^\mathrm{S}(|\mu|>10, \zs=2)$, with the former corresponding to the lens-plane solid angle over which lines of sight from the observer to the lens have $|\mu| > 10$ and the latter corresponding to the (un-lensed) solid angle in the source plane that will be magnified by $|\mu|>10$. The source plane cross-section is therefore the relevant cross-section for calculating the probability that randomly located sources at $\zs=2$ will be magnified by this lens by $|\mu|>10$.

An SIS with Einstein radius $\theta_\mathrm{E}$, has an image-plane magnification profile \citep[e.g.][]{MeneghettiLensing}
\begin{equation}
\mu_\mathrm{SIS} = \frac{\theta}{\theta-\theta_\mathrm{E}},
\label{eq:SIS_mu}
\end{equation}
where $\theta$ is the angular distance from the centre of the SIS profile. Armed with the magnification as a function of radius, we can then ask what the lens (or source) plane solid angle above a particular magnification threshold is. Note that the magnification can be both positive and negative, with negative values corresponding to images with inverted parity. In this study we are interested in the brightness of lensed point sources, for which the absolute value of $\mu$ is the relevant quantity. The image plane solid angle with $|\mu| > \mu_0$ is
\begin{equation}
\sigma_\mathrm{lens}^\mathrm{I}(>\mu_0) = \int_{\theta_0}^{\theta_1} 2 \pi \theta \, \mathrm{d}\theta 
\end{equation}
where from equation~\eqref{eq:SIS_mu} we have that $\theta_0 = \theta_\mathrm{E} \, \mu_0 / (\mu_0 + 1)$ and $\theta_1 = \theta_\mathrm{E}\, \mu_0 / (\mu_0 - 1)$, such that
\begin{equation}
\sigma_\mathrm{lens}^\mathrm{I}(>\mu_0) =   \theta_\mathrm{E}^2 \frac{4 \pi \mu_0^3}{(\mu_0^2-1)^2}.
\label{eq:Al}
\end{equation}
An infinitesimal solid angle of the image plane will map back into an infinitesimal solid angle of the source plane, which is smaller by a factor $|\mu|$ (i.e. $\mathrm{d} \sigma_\mathrm{lens}^\mathrm{I} = |\mu| \mathrm{d} \sigma_\mathrm{lens}^\mathrm{S}$), so the source plane solid angle with $|\mu| > \mu_0$ is
\begin{equation}
\sigma_\mathrm{lens}^\mathrm{S}(>\mu_0) = \int_{\theta_0}^{\theta_1} \frac{2 \pi \theta}{|\mu(\theta)|} \, \mathrm{d}\theta.
\label{eq:sigma_lens_SIS}
\end{equation} 
Note that there is a subtlety when discussing the source plane solid angle with some property (such as $|\mu| > \mu_0$), because regions of the source plane can map to multiple regions in the image plane. As equation~\eqref{eq:sigma_lens_SIS} is defined as an integral over the image plane, points on the source plane that map to multiple points on the image plane will be counted multiple times, once for each image that meets the respective property. This means that for a number density of sources per unit solid angle in the source plane, $n_\mathrm{s}$, we expect to see $\sigma_\mathrm{lens}^\mathrm{S}(>\mu_0) \times n_\mathrm{s}$ images of those sources magnified by greater than $\mu_0$ by a particular lens. This will typically be greater than the number of sources that have at least one image magnified by greater than $\mu_0$, because highly magnified lines of sight are typically multiply imaged.

Evaluating the integral in equation~\eqref{eq:sigma_lens_SIS} we find 
\begin{equation}
\sigma_\mathrm{lens}^\mathrm{S}(>\mu_0) =   \theta_\mathrm{E}^2 \frac{2 \pi (\mu_0^2 + 1)}{(\mu_0^2-1)^2}.
\label{eq:As}
\end{equation}

In equations~\ref{eq:Al} and \ref{eq:As}, $\sigma_\mathrm{lens}(>\mu_0)$ is separable into the product of $\theta_\mathrm{E}^2$ and a function of $\mu_0$. As such, the relative contribution of different haloes to the optical depth is independent of the exact definition of the optical depth, whether it is the solid angle within the Einstein radius, or the source/image plane solid angle with a magnification greater than some $\mu_0$. While different definitions of what constitutes `strong lensing' will change the total optical depth to strong lensing, it will change the optical depth of each system in the same manner, keeping their relative contributions fixed. In particular, for large magnification thresholds ($\mu_0 \gg 1$),  $\sigma_\mathrm{lens}^\mathrm{S} (>\mu_0) \propto 1/ \mu_0^2$.

While we have shown here that $P(|\mu| > \mu_0) \propto 1/ \mu_0^2$ for sources behind an SIS lens, this behaviour can be shown to be more general. Critical curves are curves in the image plane along which the magnification is  formally infinite in the geometric optics limit, while caustics are curves in the source plane found from mapping the critical curve to the source plane using the gravitational deflection angles. By Taylor expanding the gravitational potential about a point on a critical curve, it can be shown that the magnification varies inversely as the square root of the perpendicular distance of the source from the caustic \citep[e.g.][equation 35]{2002ApJ...574..970G}. The source plane solid angle above some magnification, $\mu_0$, is the length of the caustic multiplied by the distance from the caustic at which the magnification drops to $\mu_0$, $l_0$. As $l_0 \propto 1/ \mu_0^2$, the high-magnification cross-section is $\sigma_\mathrm{lens}^\mathrm{S}(> \mu_0) \propto 1/\mu_0^2$.

\subsection{The halo mass function}
\label{sect:hmf}

The halo mass function has been well studied, both in the context of analytical predictions \citep{1974ApJ...187..425P, 2001MNRAS.323....1S} and measurements from $N$-body simulations \citep[e.g.][]{2001MNRAS.321..372J, Tinker2008}. Here, we use the \citet{Tinker2008} mass function with a \citet{2014A&A...571A...1P} cosmology, as implemented in the Python library \textsc{hmf} \citep{2013A&C.....3...23M}. Throughout this paper we use $M_{200}$ to define halo masses; where $r_{200}$ is the radius at which the mean enclosed density is 200 times the critical density, and $M_{200}$ is the mass within $r_{200}$.

\begin{figure}
        \centering
        \includegraphics[width=\columnwidth]{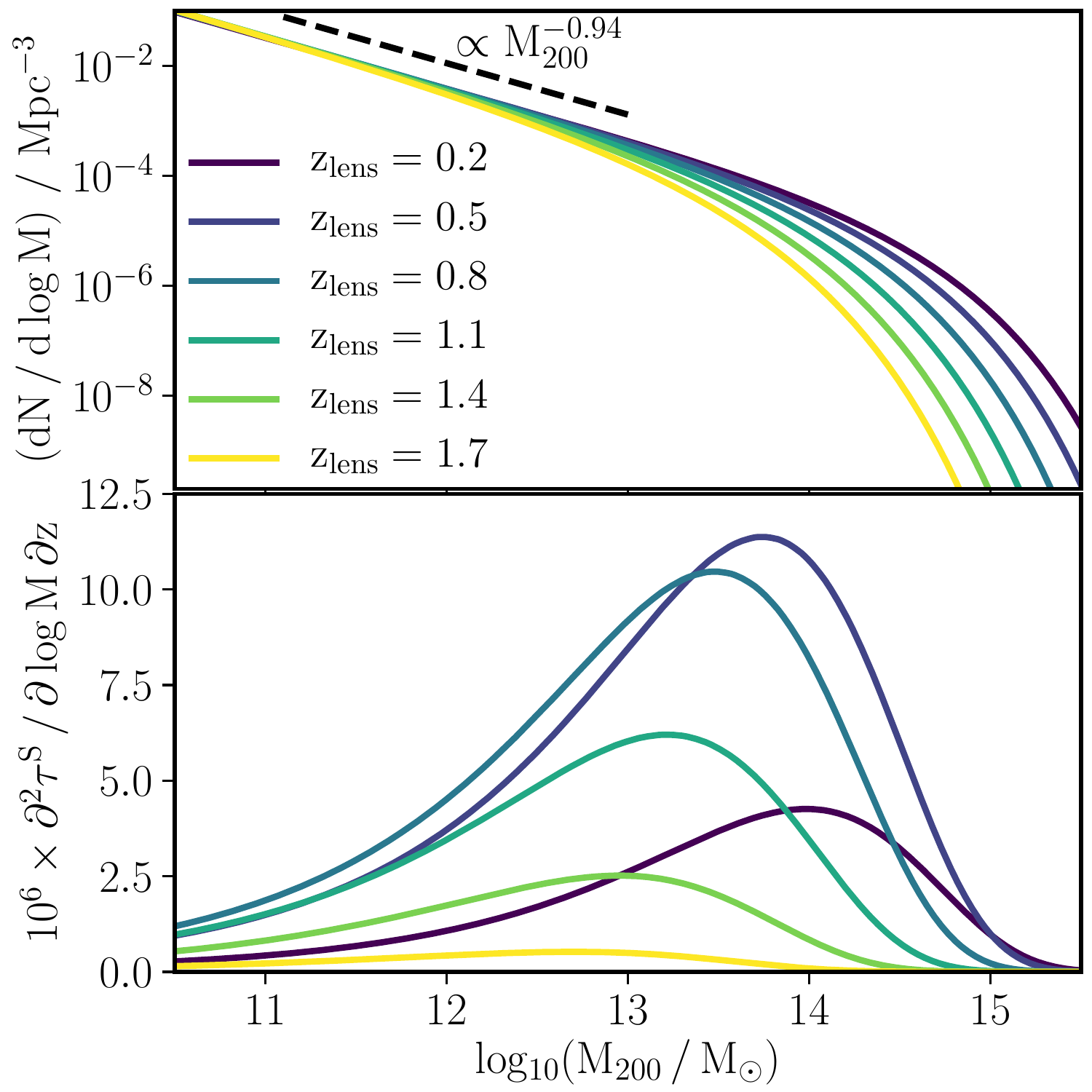}
	\caption{Top: the halo mass function at different redshifts. At all redshifts the low mass end of the mass function is approximately a power-law with a slope slightly shallower than $-1$ as indicated by the dashed line which is a power-law with a slope of $-0.94$. Bottom: the optical depth per decade in halo mass and per unit redshift, at the same redshifts as the mass function is shown in the top panel, assuming all haloes to be singular isothermal spheres. The power-law dependence of the lensing cross-section on halo mass (Section~\ref{sect:M200_SIS}) multiplied by the mass function, leads to a gently increasing power-law at low masses, which is exponentially suppressed at high masses due to the suppression of the mass function.}
	\label{fig:ODSL_SIS}
\end{figure}

We define $n(M,z)$ as the comoving number density of haloes with $M_\mathrm{200} < M$ at redshift $z$. $\partial n(M,z) / \partial \log_{10} M$ is then the number density of haloes per decade in halo mass, which is plotted for different redshifts in the top panel of Fig.~\ref{fig:ODSL_SIS}. For brevity we will often drop the `200' from halo masses, and will not explicitly write the base (always 10) of logarithms.

\subsubsection{Tying $M_{200}$ to an SIS density profile}
\label{sect:M200_SIS}

In order to relate an SIS with a given Einstein radius to a halo described by its $M_{200}$, it helps to be concrete about the normalisation of the SIS density profile. Starting with a density profile with a normalisation described by the velocity dispersion, $\sigma_\mathrm{v}$,
\begin{equation}
\rho_\mathrm{SIS} (r) = \frac{\sigma_\mathrm{v}^2}{2 \pi G r^2},
\label{eq:rho_SIS}
\end{equation}
we can integrate along a line of sight a projected radius $R$ from the centre of the halo to get the projected surface density
\begin{equation}
\Sigma_\mathrm{SIS} (R) = \frac{\sigma_\mathrm{v}^2}{2 G R}.
\label{eq:sigma_SIS}
\end{equation}
The mass enclosed within a 2D radius is then
\begin{equation}
M_\mathrm{SIS} (<R) = \frac{\pi \sigma_\mathrm{v}^2}{G}R.
\label{eq:M_SIS_2D}
\end{equation}
For a given cosmology and lensing geometry we can define the critical surface density for lensing, $\Sigma_\mathrm{crit}$, as
\begin{equation}
\Sigma_\mathrm{crit} = \frac{c^2}{4 \pi G} \frac{d_\mathrm{A}(\zs)}{d_\mathrm{A}(\zl) \, d_\mathrm{A}(\zl, \zs)}.
\label{sigma_crit}
\end{equation}
Here, $d_\mathrm{A}(\zs)$, $d_\mathrm{A}(\zl)$, and $d_\mathrm{A}(\zl, \zs)$ are the angular diameter distances between the observer and the source, the observer and the lens, and the lens and the source respectively. For an axisymmetric lens, the average surface density within the Einstein radius is $\Sigma_\mathrm{crit}$ (i.e. $M_\mathrm{SIS}(<R_\mathrm{E}) = \Sigma_\mathrm{crit} \pi R_\mathrm{E}^2$). Combining this with the fact that the physical Einstein radius is related to an angular Einstein radius by the angular diameter distance to the lens ($R_\mathrm{E} = \theta_\mathrm{E} \, d_\mathrm{A}(\zl)$) we find that
\begin{equation}
\theta_\mathrm{E} = \frac{\sigma_\mathrm{v}^2}{G \Sigma_\mathrm{crit} d_\mathrm{A}(\zl)}.
\label{eq:Re_SIS}
\end{equation}

Returning to Equation~\eqref{eq:rho_SIS}, we can integrate the density profile to find that the mass within a 3D radius is
\begin{equation}
M_\mathrm{SIS} (<r) = \frac{2 \sigma_\mathrm{v}^2}{G} r.
\label{eq:M_SIS_3D}
\end{equation}
Equating the mass within $r_{200}$ with $M_{200}$ allows us to write $\sigma_\mathrm{v}$, and hence $\theta_\mathrm{E}$, in terms of the halo mass:
\begin{equation}
\theta_\mathrm{E}(M,z_\mathrm{lens}) = \frac{M}{2 r_{200} \Sigma_\mathrm{crit} d_\mathrm{A}(\zl)}.
\label{eq:theta_E_M200}
\end{equation}
At fixed $\zl$, $\Sigma_\mathrm{crit}$ and $d_\mathrm{A}(\zl)$ are constant, and $r_{200} \propto M^{1/3}$, such that $\theta_\mathrm{E} \propto M^{2/3}$. Then, using equation~\eqref{eq:As}, $\sigma_\mathrm{lens}^\mathrm{S} \propto M^{4/3}$.

\subsection{Strong-lensing optical depths}
\label{sec:opitcal_depths}

If $\sigma_\mathrm{tot} (M, z)$ is the sum of $\sigma_\mathrm{lens}$ over all haloes with $M_{200} < M$, within some volume $V$, at redshift $z$, then
\begin{equation}
\pdv{\sigma_\mathrm{tot} (M,z)}{V}{\log M} = \pdv{n(M,z)}{\log M} \sigma_\mathrm{lens} (M,z).
\label{eq:d2sigma_tot}
\end{equation}

For a patch of sky with solid angle, $\Omega$, and with a total source-plane cross-section, $\sigma_\mathrm{tot}^\mathrm{S}$, we can define a source-plane \emph{optical depth} $\tau^\mathrm{S} \equiv \sigma_\mathrm{tot}^\mathrm{S} / \Omega$. Note that this definition of $\tau^\mathrm{S}$ differs somewhat from the source plane optical depth used by (for example) \citet{1992grle.book.....S}, because of how it treats multiply imaged regions of the source plane (see Section~\ref{sect:sigma_lens} for a discussion of this with an isolated lens). We illustrate our definition with an example, consisting of $N^\mathrm{S}$ point-source objects randomly distributed in the source plane. If we suppose that these objects must be magnified by $|\mu|>\mu_0$ in order to be detectable, then the expected number of detectable images will be $\tau^\mathrm{S}_{|\mu|>\mu_0} N^\mathrm{S}$, which can be larger than the number of different sources that are detected, due to some sources being detected multiple times. This is in contrast to the \citet{1992grle.book.....S} definition, which we call $\tilde{\tau}^\mathrm{S}_{|\mu|>\mu_0}$ following a similar discussion in \citet{2007MNRAS.382..121H}. $\tilde{\tau}^\mathrm{S}_{|\mu|>\mu_0} N^\mathrm{S}$ is the expected number of different sources that we would detect, some of which we may detect multiple times. Using $\tau^\mathrm{S}$ over $\tilde{\tau}^\mathrm{S}$ has two key advantages: firstly that it is much easier to compute, because the calculation can be done in the lens plane without having to find which regions of the lens plane map onto a common region of the source plane; and secondly because the number of images (rather than sources) is easier to count observationally.

Note that by performing our calculations exclusively in the lens plane we cannot consider complications that arise when multiple images are blended into one. There could, for example, be blended images with a total magnification $\mu_\mathrm{tot} > \mu_0$ even though neither individual image has $|\mu|>\mu_0$. This is highly unlikely for lensed gravitational waves, where the length of a `chirp' is typically less than a second, and so multiple images will only rarely overlap in time, however, it could be important for understanding other populations of lensed objects. For an SIS lens the time delay between multiple images of the same source is $\Delta t \propto \theta_E^2 / \mu_\mathrm{tot}$ \citep[e.g.][]{2002ApJ...568..488O}, and would be around an hour for an SIS with $M_{200} = 10^{11.5} \msun$, $\zl = 0.5$, $\zs = 2$,  and $\mu_\mathrm{tot}=100$. This halo mass is the lowest relevant for high magnification strong lensing (see Fig.~\ref{fig:ODSL_combo}) and so extremely large magnifications would be required for time delays of a second or less.

Defining $V(z)$ as the comoving volume out to redshift $z$, and $\Omega = 4 \pi$ as the solid angle of the whole sky, we find that
\begin{equation}
\pdv{\tau}{\log M \,}{z} = \frac{1}{4 \rm{\pi}} \pdv{\sigma_\mathrm{tot} (M,z)}{V\,}{\log M} \dv{V}{z},
\label{eq:d2tau_dz_dlogM}
\end{equation}
which using equation~\eqref{eq:d2sigma_tot} can be calculated from the halo mass function and $\sigma_\mathrm{lens} (M,z)$. 

\subsubsection{Strong-lensing optical depths in an SIS universe}

Using $\sigma_\mathrm{lens}(M,z)$ for SIS density profiles in equation~\eqref{eq:d2tau_dz_dlogM} we can calculate the relative contribution of different lens masses at different lens redshifts to the total optical depth. This is plotted (for an optical depth corresponding to $|\mu|>10$ for $\zs=2$ sources) in the bottom panel of Fig.~\ref{fig:ODSL_SIS}, where it can be seen that the relative contribution of different lens masses shifts towards higher masses at lower redshifts, peaking at $M_{200} \approx 10^{13} \msun$ at $\zl=1.4$ and $M_{200} \approx 10^{14} \msun$ at $\zl=0.2$. This is driven by the increase in the halo mass above which the mass function is exponentially suppressed (often known as $M_*$), through cosmic time. At all times there is a peak in $\partial^2 \tau / \partial \log M \, \partial z$. Lensing by low-mass haloes is relatively unimportant because the power-law slope of the mass-function ($\propto M^{-0.94}$) is not steep enough to compensate for the decreasing cross-section of low mass lenses ($\propto M^{4/3}$), while the high-mass end is suppressed by the high-mass end cut-off in the mass function.

Aside from the relative contribution to the total lensing optical depth of different lens masses, Fig.~\ref{fig:ODSL_SIS} also demonstrates the relative contribution from different lens redshifts. Of the redshifts we show, $\zl=0.5$ and $0.8$ have the largest optical depths per unit redshift (the areas under the curves in the bottom panel of Fig.~\ref{fig:ODSL_SIS}). As well as the evolution of the mass function, there are a number of different factors that contribute to this. One is the comoving volume per unit redshift, $\dv{V}{z}$, which increases with increasing redshift out to around $z=2.5$. Another is the \emph{lensing efficiency}, the inverse of the critical surface density, which at fixed source redshift is proportional to $d_\mathrm{A}(\zl) d_\mathrm{A}(\zl, \zs)$. In a flat and non-expanding universe, $d_\mathrm{A}(\zl) + d_\mathrm{A}(\zl, \zs) = d_\mathrm{A}(\zs)$. At fixed $d_\mathrm{A}(\zs)$, the product of $d_\mathrm{A}(\zl)$ and $d_\mathrm{A}(\zl, \zs)$ is then maximised when they are equal, i.e. when the lens is halfway between the observer and source. This is complicated somewhat by a universe that is expanding, but it remains true that lensing is most efficient when the lens is neither close to the observer nor close to the source, but roughly midway between. In the case of a $\zs = 2$ source, and a \citet{2014A&A...571A...1P} cosmology, the lensing efficiency peaks for $\zl=0.52$.

\section{Lensing in hydrodynamical simulations}
\label{sect:lensing_in_hydro}

Cosmological hydrodynamical simulations have increased dramatically in their realism over the past decade and can now reproduce many of the key properties of observed galaxies \citep[e.g.][]{2014Natur.509..177V, 2015MNRAS.446..521S, 2016MNRAS.462.3265D, 2017MNRAS.467.4739K, 2018MNRAS.473.4077P, 2019MNRAS.486.2827D} as well as galaxy groups and clusters \citep[e.g.][]{2013MNRAS.429..323S, 2014MNRAS.441.1270L, 2017MNRAS.465..213B, 2017MNRAS.465.2936M, 2017MNRAS.470.4186B, 2017MNRAS.471.1088B, 2018MNRAS.475..648P, 2018MNRAS.480.2898C}. Importantly for lensing, simulations can produce populations of galaxies with the correct distribution of stellar mass, and the correct stellar mass -- size relation \citep{2015MNRAS.446..521S}. As such, these simulations can now be used to answer the question of how important different objects are as gravitational lenses, from a theoretical perspective. 

In order to determine the contribution of different lens masses to the strong lensing optical depth as predicted from a state of the art hydrodynamical simulation, we require the simulation to resolve the strong lensing region of the lowest mass haloes important for lensing ($\sim \num{3e11} \msun$, see Fig.~\ref{fig:ODSL_combo}), while simultaneously covering a large enough volume to accurately sample the high-mass end of the halo mass function. A single hydrodynamical simulation like this does not currently exist, so instead we combine two simulations, using the high-resolution publicly available\footnote{The galaxy and halo catalogues of the simulation suite, as well as the particle data, are publicly available at \href{http://www.eaglesim.org/database.php}{http://www.eaglesim.org/database.php} \citep{2016A&C....15...72M, 2017arXiv170609899T}} $(100 \mpc)^3$ \eagle simulation \citep{2015MNRAS.446..521S, 2015MNRAS.450.1937C} to resolve low mass haloes, and the large volume of the $(400 \, h^{-1} \mpc)^3$ \bahamas simulation \citep{2017MNRAS.465.2936M} to have an adequate number of high mass haloes.

\subsection{Simulation descriptions}

\eagle and \bahamas have similar sub-grid physics models for galaxy formation physics on scales below the resolution limits of the simulations. These include models for gas cooling, star formation, and feedback both from stars and active galactic nuclei (AGN). \eagle was run using a \citet{2014A&A...571A...1P} cosmology,\footnote{With $\Omega_\mathrm{m}=0.307$, $\Omega_\mathrm{b}=0.04825$, $\Omega_\mathrm{\Lambda}=0.693$, $\sigma_8 = 0.8288$, $n_\mathrm{s} = 0.9611$ and $h = 0.6777$.} and for \bahamas (which has been run with many different cosmologies, with and without massive neutrinos) we used the simulation based on the same \citet{2014A&A...571A...1P} results as \eagle's cosmology, with zero neutrino mass, which is described in \citet{2018MNRAS.476.2999M}.\footnote{This differs slightly from the \eagle cosmology, with $\Omega_\mathrm{m}=0.3175$, $\Omega_\mathrm{b}=0.049$, $\Omega_\mathrm{\Lambda}=0.6825$, $\sigma_8 = 0.8341$, $n_\mathrm{s} = 0.9624$ and $h = 0.6711$. We assume the \eagle cosmology for all of our lensing calculations, including those with \bahamas.}

For \eagle, the DM and initial baryon particle masses are $\num{9.7e6} \msun$ and $\num{1.8e6} \msun$ respectively, while for \bahamas they are $\num{6.6e9} \msun$ and $\num{1.2e9} \msun$. The corresponding Plummer-equivalent gravitational softening lengths are $0.7 \kpc$ for \eagle and $6.0 \kpc$ for \bahamas, with these being fixed physical lengths (i.e. not comoving) at all redshifts we consider.

\subsection{Lensing description}

Our lensing procedure treats each halo as an isolated lens, ignoring the effect of other structures along the line-of-sight. This is done because we want to assign light-rays that have high magnifications to a single lensing object, in order to answer the question of what was responsible for the lensing. Ignoring the effects of multiple lens planes is justified by the fact that only a very small fraction of light-rays meet our criterion for being strongly lensed, such that the probability of an object being sufficiently aligned to be strongly lensed by two separate haloes is negligible. This was verified explicitly by \citet{2007MNRAS.382..121H}, who did full multi-plane ray-tracing through the Millennium simulation \citep{2005Natur.435..629S}, and found that strong lensing events can almost always be traced to a single dominant lensing object. We stress that this does not mean that line-of-sight structures can be ignored in detailed lens modelling of individual systems, where including line-of-sight structures in the model can improve the match between the predicted and observed positions of multiply imaged sources \citep[for example][]{2018A&A...614A...8C}. It means that the lensing cross-sections of lenses are not significantly altered on average by objects along the line-of-sight.   

Our procedure for generating lensing maps from simulated haloes follows \citet{2019MNRAS.488.3646R}, who studied the Einstein radii of galaxy clusters from \bahamas simulations run with different DM models. For each halo at each snapshot redshift, we first find all mass within $5 \, r_{200}$ of the particle with the lowest gravitational potential energy. We then calculate the projected surface density, $\Sigma$, on a regular grid using an SPH-like smoothing scheme based on the distance to the 16th nearest neighbour. We make 3 maps of each halo -- projecting the mass along 3 orthogonal lines of sight, for which we use the simulation $x$, $y$ and $z$ axes. These maps are square, with a side-length of $2 \, r_{200}$, and with $1024 \times 1024$ pixels. The resolution therefore increases in lower mass haloes, where the critical curve is on a smaller physical scale. At $z=0$ this corresponds to a pixel scale of $4.5 \kpc$ for a $10^{15} \msun$ halo and $0.2 \kpc$ for a $10^{11} \msun$ halo. Note that the numerical values quoted here were motivated by the convergence tests discussed in Section~\ref{sec:conv}.

\begin{figure}
\begin{center}
\begin{tikzpicture}[scale=0.3]

\clip (-13,-1) rectangle + (29,30);

\tikzmath{\ha = 20; \zmax=0.4; \zs0=0; \zs1=0.125; \zs2=0.25; \zs3=0.375; \zi0=0.5*(\zs0+\zs1);} 
\draw [fill=red!10,solid] (90-\ha:0)--(90-\ha:6.25) arc (90-\ha:90+\ha:6.25) -- (90+\ha:6.25) arc (90+\ha:90-\ha:0);
\draw [fill=green!10,solid] (90-\ha:6.25)--(90-\ha:17.725) arc (90-\ha:90+\ha:17.725) -- (90+\ha:6.25) arc (90+\ha:90-\ha:6.25);
\draw [fill=blue!10,solid] (90-\ha:17.725)--(90-\ha:31.25) arc (90-\ha:90+\ha:31.25) -- (90+\ha:17.725) arc (90+\ha:90-\ha:17.725);

\draw [dashed] (90-\ha:12.5) arc (90-\ha:90+\ha:12.5);
\draw [dashed] (90-\ha:25) arc (90-\ha:90+\ha:25);

\node [right] at (90-\ha:0) {\Large $z^\mathrm{snap}=0$};
\node [right] at (90-\ha:12.5) {\Large $z^\mathrm{snap}=0.125$};
\node [right] at (90-\ha:25) {\Large $z^\mathrm{snap}=0.25$};

\foreach \z in {1,...,10}
{ \draw [solid] (90+\ha:4*\z-2) arc (90+\ha:90+\ha+8/\z:4*\z-2);
\node [left] at (90+\ha+5/\z:4*\z-2) {\Large $z^\mathrm{L}_{\z}$};
}

\draw [blue, pattern=north east lines, pattern color=blue!40] (90-\ha:20)--(90-\ha:24) arc (90-\ha:90+\ha:24) -- (90+\ha:20) arc (90+\ha:90-\ha:20);
\draw [<->,shift={(+0.5,-0.2)}] (90-\ha:20) --node[anchor=west,rotate=0]{\Large$\delta z$} (90-\ha:24);
\draw [blue,solid] (90+\ha:22) arc (90+\ha:90+\ha+8/6:22);
\node [blue] at (90:22) {\LARGE$\delta V_6$};

\draw [red, pattern=north east lines, pattern color=red!40] (90-\ha:4)--(90-\ha:8) arc (90-\ha:90+\ha:8) -- (90+\ha:4) arc (90+\ha:90-\ha:4);
\draw [<->,shift={(+0.5,-0.2)}] (90-\ha:4) --node[anchor=west,rotate=0]{\Large$\delta z$} (90-\ha:8);
\draw [red,solid] (90+\ha:6) arc (90+\ha:90+\ha+4:6);
\node [red] at (90:6) {\LARGE $\delta V_2$};

\draw [green, pattern=north east lines, pattern color=green!40] (90-\ha:8.1)--(90-\ha:12) arc (90-\ha:90+\ha:12) -- (90+\ha:8.1) arc (90+\ha:90-\ha:8.1);
\draw [<->,shift={(+0.5,-0.2)}] (90-\ha:8.1) --node[anchor=west,rotate=0]{\Large$\delta z$} (90-\ha:12);
\draw [green,solid] (90+\ha:10) arc (90+\ha:90+\ha+2.4:10);
\node [green] at (90:10) {\LARGE $\delta V_3$};

\draw [solid] (0,0) -- (90-\ha:100 * \zmax) (0,0) -- (90+\ha:100 * \zmax) ;

\node [above] at (0,1.1) {\Large $\Omega$};
\draw [solid,style=double] (90+\ha:2.7) arc (90+\ha:90-\ha:2.7);

\end{tikzpicture}
\caption{A schematic illustration of our volume of lenses, using the snapshot redshifts from \bahamas. The redshift range shown is covered by the three lowest redshift snapshots from \bahamas, with the red, green and blue filled regions covering redshifts closest to $z^\mathrm{snap} = 0, 0.125$ and $0.25$ respectively. The ticks along the left edge of the wedge show the redshifts at which we evaluate the contribution to the strong-lensing optical depth, which is done by taking the comoving volume corresponding to a redshift interval $\delta z$. When this redshift interval includes regions closer in redshift to two different snapshots (i.e for $z^\mathrm{L}_2$) the optical depth calculation uses only the snapshot closest in redshift to the centre of this redshift interval.}
\label{fig:lightcone}
\end{center}
\end{figure}
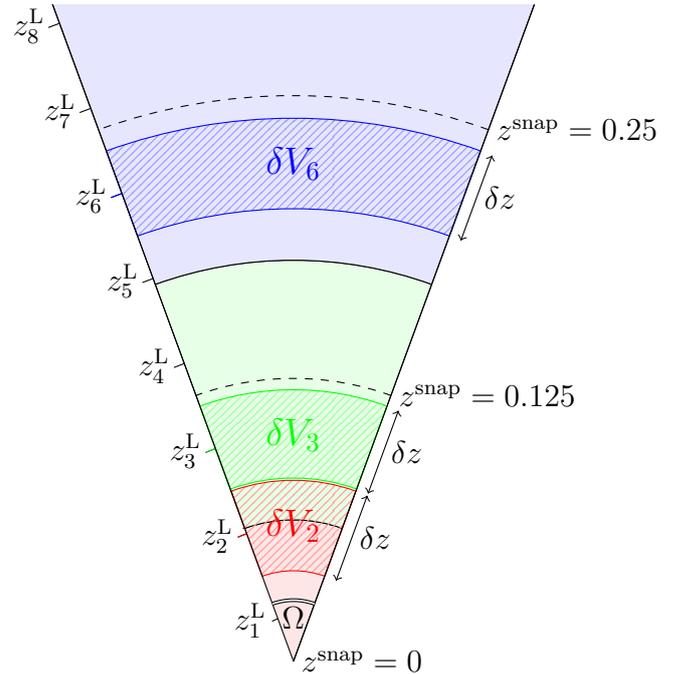

Our calculation of $\partial^2 \tau / \partial \log M \, \partial z$ is done at more lens redshifts than we have simulation snapshots, as illustrated in Fig.~\ref{fig:lightcone}. At each redshift that we use as a lens plane, $\zl$, we find the snapshot closest to it in redshift. We then use the $\Sigma$ maps of the haloes from this snapshot to calculate the contribution to the strong lensing optical depth from this particular $\zl$. This is done by first dividing $\Sigma$ by $\Sigma_\mathrm{crit}$, to get the dimensionless convergence, $\kappa$. As both $\kappa$, and the gravitational shear, $\vect{\gamma}$, are second derivatives of the projected Newtonian potential, they can be readily calculated from one another using discrete Fourier transforms \citep[see e.g.][]{2019MNRAS.488.3646R}. We can then make a map of the magnification $\mu$ using
\begin{equation}
\mu = \frac{1}{(1- \kappa)^2 - |\vect{\gamma}|^2}.
\label{eq:mag_map}
\end{equation}

For lens planes at a different redshift from the snapshot used, we keep $\Sigma$ as a function of physical coordinates fixed. For different lens planes that use the same snapshot, the differences between them are that $\Sigma_\mathrm{crit}$ changes with $\zl$ (leading to a difference in the relationship between $\Sigma$ and the convergence, $\kappa$) as does the relationship between physical distances in the lens plane and angles on the sky.

\subsection{Cross-sections and optical depths}

With $\mu$ on a regular grid in the image plane, we calculate the source plane solid angle with $|\mu| > \mu_0$ behind some lens as
\begin{equation}
\sigma_\mathrm{lens}^\mathrm{S} = \sum_{|\mu| > \mu_0} \frac{\sigma_\mathrm{pixel}}{|\mu|} ,
\end{equation}
where $\sigma_\mathrm{pixel}$ is the solid angle of each image-plane pixel, and the sum is over all pixels with ${|\mu| > \mu_0}$.\footnote{We stress again that this is not strictly the source plane solid angle that is magnified by greater than $\mu_0$, as it counts multiply imaged regions of the source plane multiple times.} 

At each lens plane redshift ($z^\mathrm{L}_i = 0.04i - 0.02$ for $i = 1, 2, ... 50$), we calculate $\sigma_\mathrm{lens}^\mathrm{S}$ of each halo with $M_{200} > 10^{11} \msun$ in \eagle and $M_{200} > 10^{13.5} \msun$ in \bahamas. We bin haloes by $\log_{10} (M_{200} / \msun)$ and sum up $\sigma_\mathrm{lens}$ within each bin, where the bin width is $\Delta \log_{10} M_{200} = 0.18$. Dividing this sum by three times\footnote{To reflect the three projection axes.} the simulation volume and the log-mass bin width we get the left side of equation~\eqref{eq:d2sigma_tot}, which we can convert to $\partial^2 \tau / \partial \log M \, \partial z$ using equation~\eqref{eq:d2tau_dz_dlogM}. 

\subsection{Numerical convergence of strong lensing cross-sections}
\label{sec:conv}

Before we did our full lensing analysis of the \eagle and \bahamas simulations, we first performed a number of tests to determine that our lensing procedure produced numerically converged results. We did these tests on a single snapshot from each of \eagle  and \bahamas, so that we could test a large number of possible numerical parameters. These snapshots were chosen to be at a redshift that contributes significantly to the lensing of high redshift sources, and to be at a redshift where \eagle and \bahamas have a similarly timed output. We chose to use the $z=0.366$ snapshot from \eagle, and the $z=0.375$ snapshot from \bahamas.

Accurately calculating the distribution of magnifications due to some mass distribution requires that the critical curves be adequately captured. This in turn requires that the mass distribution within the critical curves is sampled with a reasonable number of pixels. Smaller haloes have smaller critical curves, and hence require higher resolutions, while larger haloes need large fields of view to include the full halo's mass distribution. In order to achieve both of these, we used
 a pixel size and field of view that both increase with increasing halo mass, specifically in proportion to the virial radius ($\propto M_{200}^{1/3}$). 

As both our pixel size and field of view scale with the virial radius, the number of pixels used for the mass maps is independent of halo mass. We experimented with 256, 512, 1024 and 2048 pixels on a side, making square maps with a side-length of $2 r_{200}$. The pixel scale acts as a scale on which the mass distribution is smoothed, and with the largest pixels ($r_{200} / 128$), our lensing cross-sections were substantially reduced compared to smaller pixel cases. The two smallest pixel scales ($r_{200} / 512$ and $r_{200} / 1024$) produced converged results, suggesting that the larger of those two pixel scales (corresponding to 1024 pixels on a side) is more than adequate.

The reason that our lensing procedure becomes insensitive to the pixel scale is because it already smooths out the mass of individual particles on a scale that depends on the local number density of particles. This means that so long as the pixels are not too large, smaller pixels do not lead to smaller structures being resolved, but rather just a finer sampling of a smooth mass distribution. Because the resolution of our lensing maps is set by the density of simulation particles, we needed to check that a different simulation resolution, with a different number density of simulation particles, would have provided converged results.

To investigate this, we made lensing maps of all haloes in our test snapshots, both with the full simulation data, and when only using a fraction, $f_\mathrm{sub}$, of the simulation particles. When making these \emph{subsampled} maps, the mass of each particle was increased by $1/f_\mathrm{sub}$ to create a lower resolution version of the same simulated mass distribution. With these subsampled versions, we could then find the halo mass at which the lensing properties of the subsampled haloes disagreed with those of the full haloes, which indicates the halo mass down to which we can trust the lensing maps being generated from full simulation data.

\begin{figure}
        \centering
        \includegraphics[width=\columnwidth]{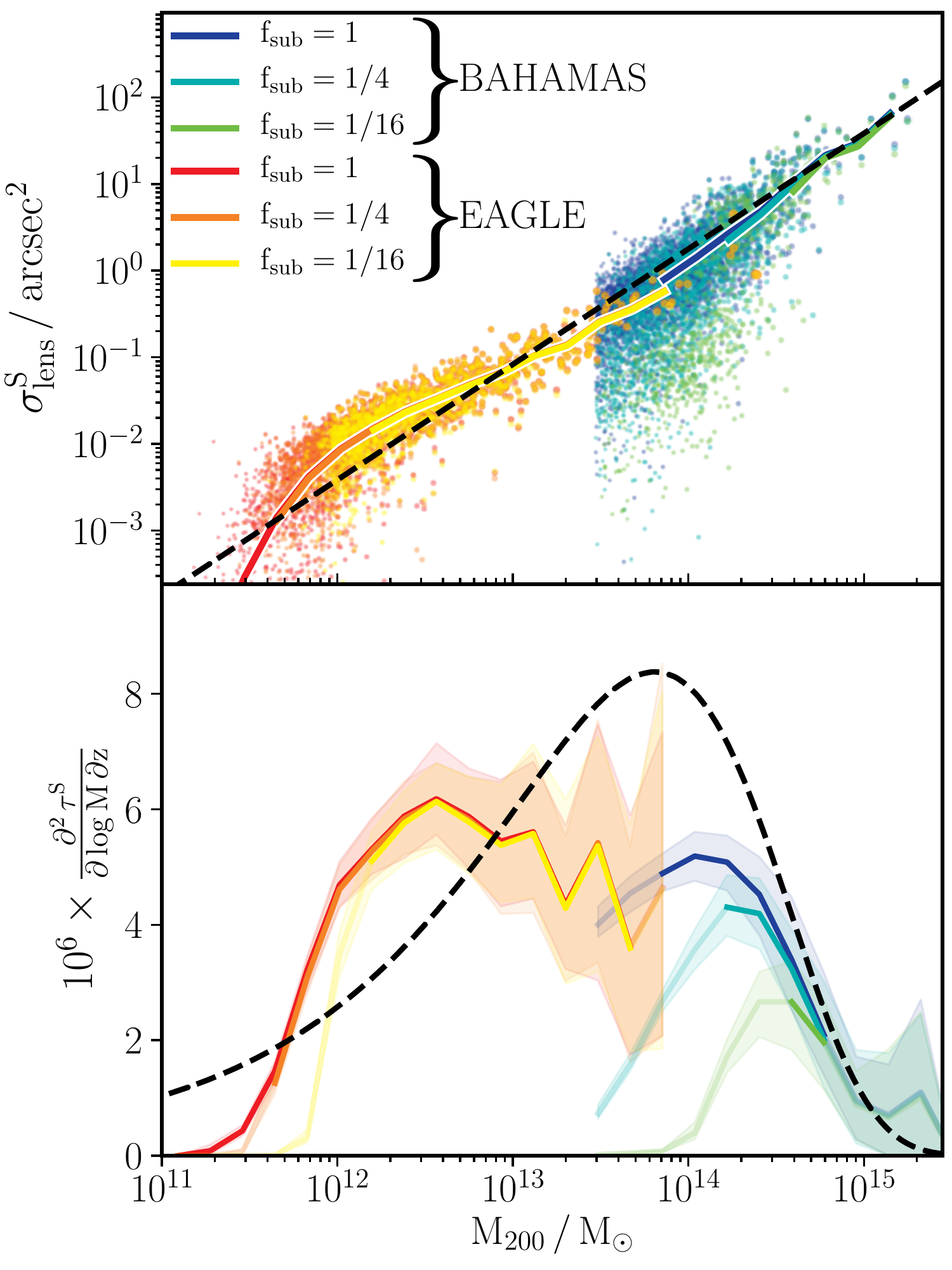}
	\caption{Top: the $\zs = 2$, $|\mu| > 10$, source-plane cross-sections for individual lenses from \eagle and \bahamas, from the respective snapshots with $z \approx 0.37$. The different colours correspond to different $f_\mathrm{sub}$ values as described in Section~\ref{sec:conv}. Running medians as a function of halo mass are shown as solid lines, which are drawn only in the mass range where the relevant simulation with the particular $f_\mathrm{sub}$ value is `converged'. The black dashed line shows the prediction for haloes modelled as singular isothermal spheres all the way out to their virial radii. Bottom: the optical depth per decade in halo mass and per unit redshift, calculated from the cross-sections in the top panel. Lines are shown as solid in the halo mass ranges where they are converged, and are faded at lower masses. We stop showing \eagle lines at the high-mass end due to the poor sampling of the mass function there. The shaded regions around the lines show the 2.5th to 97.5th percentiles when the optical depth is calculated from a bootstrap resampling of all haloes in the top panel. For the black dashed line a \citet{Tinker2008} mass function was assumed.}
	\label{fig:convergence}
\end{figure}

In the top panel of Fig.~\ref{fig:convergence} we show the $\zs = 2$ source plane cross-sections for $|\mu| > 10$, for individual haloes from both \eagle and \bahamas, using subsampling factors of $1$, $1/4$ and $1/16$. The bottom panel then shows $\partial^2 \tau / \partial \log M \,  \partial z$ with these different subsampling factors. For the case of \eagle we see that other than a slight suppression at masses below $4 \times 10^{11} \msun$, the $f_\mathrm{sub} = 1/4$ optical depth is indistinguishable from the full simulation case. We can also see that \eagle has just enough resolution to resolve the lowest mass haloes important for lensing, as had the mass resolution been 16 times worse (corresponding to the yellow line), the low-$M_{200}$ cut off in $\partial^2 \tau / \partial \log M \, \partial z$ would have been numerical rather than properly resolved.

The mass-scale at which convergence is achieved in \bahamas is, unsurprisingly, different, given the much poorer resolution of the \bahamas simulations. Defining \emph{convergence} as an agreement on $\partial^2 \tau /\partial \log M \,  \partial z$ better than 20\% between subsequent $f_\mathrm{sub}$ levels , $f_\mathrm{sub}=1/16$ is converged down to $M_{200} \approx \num{3.2e14} \msun$ and $f_\mathrm{sub}=1/4$ down to $M_{200} \approx \num{1.6e14} \msun$. Assuming a similar fractional improvement in the mass down to which we are converged going from $f_\mathrm{sub}=1/4$ to $f_\mathrm{sub}=1$ as we had when going from $f_\mathrm{sub}=1/16$ to $f_\mathrm{sub}=1/4$, we expect that the full \bahamas results are converged down to a halo mass of around $10^{14} \msun$. For this reason we use \bahamas to make predictions for the contribution of $M_{200} > 10^{14} \msun$ lenses to the strong lensing optical depth, using \eagle for halo masses below this. Owing to its relatively small box size, \eagle has few haloes at masses $10^{13.5} - 10^{14} \msun$. As we will see in the next section, this leads to our lensing calculation being most uncertain at these intermediate masses, where \bahamas is not well resolved, but \eagle suffers from a small volume.

\section{Results}
\label{sect:results}

\begin{figure*}
        \centering
        \includegraphics[width=\textwidth]{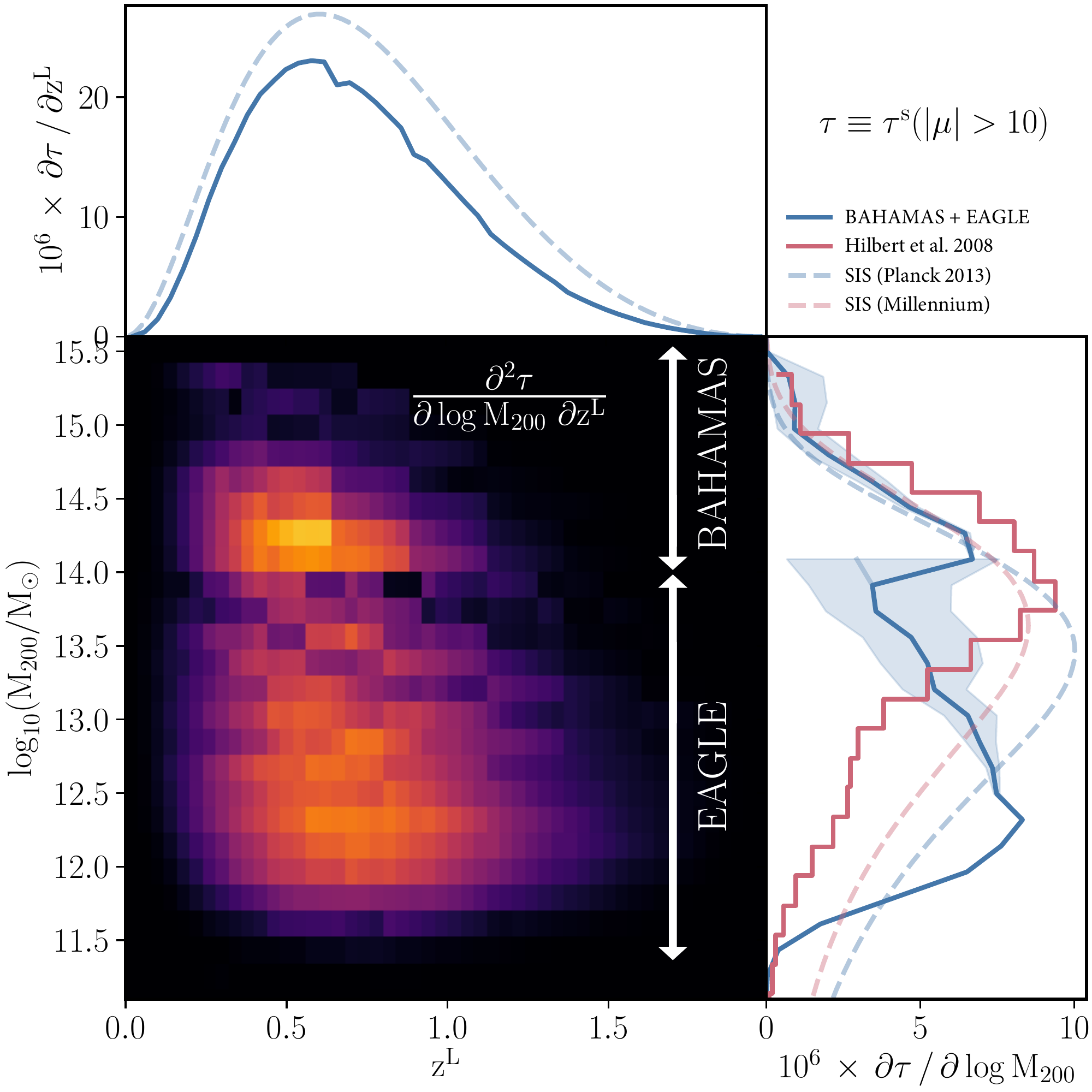}
	\caption{The main panel shows the source-plane optical depth for lensing by a magnification greater than 10, at a source redshift of 2, per unit lens redshift and per decade in lens halo mass. At halo masses above $10^{14} \msun$ we use the \bahamas simulation, while \eagle is used at lower masses. Above and to the right of the main panel we show this quantity marginalised over halo mass and lens redshift respectively as solid blue lines. For comparison, we show these same marginalised quantities for our SIS + \citet{Tinker2008} mass function model as blue dashed lines. We also show $\partial \tau / \partial \log M$ from \citetalias{Hilbert2008} as the red stepped line. \citetalias{Hilbert2008} used a different cosmology and mass definition from us, and to show the contribution of these different definitions to the differences between our results and those from \citetalias{Hilbert2008}, we show an SIS model with the cosmology and mass definition used by  \citetalias{Hilbert2008} as the red dashed line. The dip in the \bahamas + \eagle results just below $10^{14} \msun$ is where \bahamas would not have sufficient resolution (so is not being used), but \eagle's volume leads to a poorly sampled mass function.}
	\label{fig:ODSL_combo}
\end{figure*}

The differential optical depth for strong lensing, $\partial^2 \tau / \partial \log M \, \partial z$, is plotted as a function of lens redshift and halo mass in Fig.~\ref{fig:ODSL_combo}, where we calculate this quantity from \bahamas for $M_{200} > 10^{14} \msun$ and from \eagle for $M_{200} < 10^{14} \msun$. We remind the reader that this is a source plane optical depth for magnifications greater than 10. We also plot $\partial \tau / \partial z$ (from integrating over lens mass) and  $\partial \tau / \partial \log M$ (from integrating over lens redshift).

The first comment to make is that both $\partial \tau / \partial z$ and  $\partial \tau / \partial \log M$ are quite broad, so there is no one population of lenses that dominates the optical depth for high magnification lensing. Integrating over both lens mass and lens redshift, the optical depth for $|\mu|>10$ with $\zs = 2$ is approximately $\num{2e-5}$. As explained earlier, for high magnification thresholds, the optical depth is inversely proportional to the square of the threshold, such that the high magnifications required to significantly alter the true masses of compact binaries that have been detected with gravitational waves ($\mu \gtrsim 50$) will happen to only a very small fraction of all $z \sim 2$ compact binary coalescences (fewer than one in a million).

Results from the SIS model are also included in Fig.~\ref{fig:ODSL_combo}, where it can be seen that it does a reasonable job of reproducing the simulation-derived optical depth, including the relative contribution from different lens redshifts ($\partial \tau / \partial z$). This suggests that it will capture the dependence of $\tau$ on source redshift, which is the quantity required for calculating the expected number of lensed gravitational wave events that we should detect. Where this SIS model works less well is in the relative importance of different lens masses, with it predicting that low halo masses ($< 10^{12} \msun$) make a reasonable contribution to the optical depth, while the hydrodynamical simulations have a steep drop in  $\partial \tau / \partial \log M$ below $10^{12} \msun$. The mass scale at which the simulated systems become inefficient lenses is related to the stellar to halo mass relation for galaxies, whose behaviour changes at halo masses of around $10^{12} \msun$. We discuss this further in Section~\ref{sect:low_mass_end}.

As another comparison, in Fig.~\ref{fig:ODSL_combo} we plot $\partial \tau / \partial \log M$ from \citetalias{Hilbert2008}, who also used $\zs = 2$ and $|\mu|>10$ (private communication). In \citetalias{Hilbert2008} the optical depth was calculated from ray-tracing through a DM-only simulation, with the addition of analytic gravitational potentials associated with galaxies, where the mass distribution within the galaxies was taken from the results of a semi-analytic galaxy formation model. The agreement between our $\partial \tau / \partial \log M$ and that from \citetalias{Hilbert2008} is fairly good in general, although there are differences that we discuss further in Section~\ref{sect:comp_H08}.

\subsection{Random error on $\partial \tau / \partial \log M$}
 
The discontinuity in $\partial \tau / \partial \log M$ at $10^{14} \msun$ is because we change the simulation used at this mass. The discrepancy between the two simulations at this mass scale, with \eagle predicting a factor of two less lensing from $M_{200} \approx 10^{14} \msun$ haloes than \bahamas, could be for a number of reasons. Given that poor resolution leads to a decrease in the strong lensing cross-section of haloes (Fig.~\ref{fig:convergence}), it is unlikely to be resolution related. However, it could be that the different baryonic physics prescriptions lead to different predictions for the distribution of mass within haloes at this mass scale. Alternatively, it could just be random error associated with the \eagle prediction, because the low volume of \eagle means that the mass function at the high-mass end is poorly sampled. This can lead to noise both on the number of haloes and their mean lensing cross-section. The top panel of Fig.~\ref{fig:convergence} shows that the median $\sigma_\mathrm{lens}^\mathrm{S}$ of \eagle galaxies at this mass scale is a factor of almost two lower than in \bahamas, which explains the discrepancy in $\partial \tau / \partial \log M$. However, without more \eagle haloes at this mass scale, we cannot say whether this is a systematic difference, or just a random quirk of the particular sample of massive haloes in \eagle.

Using either the measured mass function from our simulations, or a mass function such as that from \citet{Tinker2008} combined with the volume of our simulations, it is possible to work out the expected number of haloes at each redshift in a given mass bin. However, using this with an assumption of Poisson statistics for the number of haloes in a given mass and redshift bin would underestimate the random error on $\partial \tau / \partial \log M$ for two reasons. Firstly, the different snapshots are not independent, because it is the same haloes (evolving through time) that appear in the different snapshots. Secondly, there is fairly large scatter in $\sigma_\mathrm{lens}^\mathrm{S}$ at fixed lens redshift and halo mass. For \bahamas, the distribution of $\sigma^\mathrm{S}_\mathrm{lens}$ at $M_{200} \approx 10^{14} \msun$, $z^\mathrm{L} = 0.375$ is well fit by a log-normal distribution with a standard deviation of 0.35 dex (this scatter can be seen in the top panel of Fig.~\ref{fig:convergence}). Such a distribution has 50\% of the signal coming from only 20\% of the objects, so that the scatter in the total lensing signal from a mass bin is significantly larger than the Poisson expectation.

In order to get an estimate of the random error on $\partial \tau / \partial \log M$ that reflects the points above, we use a bootstrap technique. Separately for both \eagle and \bahamas, we extract the 100 most massive haloes from the $z=0$ snapshot and find their primary progenitors in the preceding snapshots. We then draw 100 haloes with replacement from these 100 $z=0$ haloes, and calculate $\partial \tau / \partial \log M$ with the 100 most massive haloes and their progenitors replaced by the bootstrapped sample of 100 haloes and their progenitors. We do this 1,000 times and find the 2.5\% and 97.5\% percentiles for $\partial \tau / \partial \log M$, which are shown by the blue shaded regions in Fig.~\ref{fig:ODSL_combo}. Note that the choice of the 100 most massive haloes is fairly arbitrary. In principle we could bootstrap sample all haloes, but we avoided this because our simple method for finding progenitors (using the most massive halo with a comoving centre within $4 r_{200}$ of the halo comoving centre at the preceding snapshot) works only for the most massive haloes in the box. Owing to the sharply rising mass function towards low halo masses, the random error associated with $\partial \tau / \partial \log M$ at low masses should be negligible. Taking the noisy sampling of the high-mass end into account, the \bahamas and \eagle $\partial \tau / \partial \log M$ at $M_{200} \sim 10^{14} \msun$ are consistent at the $2\sigma$ level.

\subsection{The lowest halo masses that are efficient lenses}
\label{sect:low_mass_end}

An obvious feature in Fig.~\ref{fig:ODSL_combo} is that $\partial \tau / \partial \log M$ drops rapidly at halo masses below $10^{12} \msun$. We attribute this to the rapid fall off in stellar mass below this halo mass. Using abundance matching, the observed stellar mass function can be combined with $N$-body simulations to predict the stellar mass that resides in different halo masses at different redshifts \citep{2013ApJ...762L..31B, 2013MNRAS.428.3121M}. The resulting stellar-to-halo mass relation is matched reasonably well by the \eagle simulations \citep{2015MNRAS.446..521S}. At a given redshift, the relationship between stellar mass and halo mass is approximately a double power-law, which breaks at a characteristic halo mass of $M_1 = 10^{11.5-12} \msun$. At halo masses below this, the stellar mass ($M^*$) falls off rapidly with decreasing halo mass. In the local Universe, this fall off is approximately $M^* \propto M_{200}^{5/2}$, while at $z=2$ it is closer to $M^* \propto M_{200}^{2}$. This fall off in stellar mass at low halo masses means that stars quickly become unimportant for lensing when $M_{200} < M_1$. These low mass haloes then only have a DM component, which should be reasonably well fit by a Navarro, Frenk and White (NFW) density profile \citep{1997ApJ...490..493N}.  The strong lensing cross-section of an NFW halo decreases exponentially with decreasing halo mass below $10^{13} \msun$ \citep{2007MNRAS.382..121H}, so that these low mass haloes have very low lensing cross-sections, which -- even combined with their large abundance -- leads to them making a negligible contribution to the total optical depth.

While the median $\sigma_\mathrm{lens}^\mathrm{S}$ is a monotonically increasing function of $M_{200}$, inspection of Fig.~\ref{fig:convergence} reveals a feature around $10^{12} \msun$, where lensing is most efficient compared with the SIS prediction. This is the same mass scale at which 
$M^{*} / M_{200}$ peaks \citep{2013MNRAS.428.3121M}, and this feature in $\sigma_\mathrm{lens}^\mathrm{S} (M_{200})$ can be ascribed to the turn over of $M^{*} / M_{200}$.

\subsection{Comparison with \citet{Hilbert2008}}
\label{sect:comp_H08}

Compared with the \citetalias{Hilbert2008} $\partial \tau / \partial \log M$, hydrodynamical simulations predict slightly more strong lensing in total, with the increase primarily due to galaxies with halo masses of $10^{12} - 10^{13} \msun$. The cosmology and mass definition\footnote{\citetalias{Hilbert2008} use $M_{200,\mathrm{mean}}$ whereas we use $M_{200,\mathrm{crit}}$.} are different between \citetalias{Hilbert2008} and the hydrodynamical simulations. To investigate whether this explains the different lensing predictions, we calculated two different $\partial \tau / \partial \log M$ from our SIS model -- one using the cosmology and mass definition from \eagle and \bahamas, the other those adopted by \citetalias{Hilbert2008}. These two $\partial \tau / \partial \log M$ are plotted as dashed lines in the right panel of Fig.~\ref{fig:ODSL_combo}, and the differences between them are smaller than the differences between the lines from the different simulations.

For haloes with $M_{200} > 10^{14} \msun$, i.e. those from \bahamas, $\partial \tau / \partial \log M$ is similar between the hydrodynamical simulations and \citetalias{Hilbert2008}. \citet{2019MNRAS.488.3646R} looked at the density profiles of \bahamas clusters, both DM-only and including baryons, and found that the total density at the centre of clusters increases when simulations include baryons, not just because of the contribution from stars (which should be captured by \citetalias{Hilbert2008} who included analytical potentials associated with a stellar disc and bulge), but also because the DM profile itself becomes more centrally concentrated due to adiabatic contraction \citep{2004ApJ...616...16G}. The fact that this does not push the \bahamas curve above that from \citetalias{Hilbert2008} probably reflects the fact that in \citetalias{Hilbert2008} the analytical baryonic potential is added to the DM distribution from a DM-only simulation. Such a simulation does not ignore the baryonic material in the universe, rather it simulates it as if it were also made of DM. As such, there is more DM in a DM-only simulation than in a hydrodynamical simulation, such that when \citetalias{Hilbert2008} add in a stellar component there is now more total mass in each halo. This appears to mimic the effects of adiabatic contraction on strong lensing cross-sections such that the hydrodynamical and DM-only plus analytical galaxies predictions are similar in the galaxy cluster regime.

The largest discrepancy between our results and those from \citetalias{Hilbert2008} is in the $10^{12} - 10^{13} \msun$ mass range. At these halo masses the DM-only prediction is for negligible amounts of strong lensing \citep{2007MNRAS.382..121H}, so the lensing signal for these systems is dominated by the stellar component. In \eagle the lensing cross-sections of these galaxies are well resolved (Fig.~\ref{fig:convergence}), and in \citetalias{Hilbert2008} they use analytic expressions to calculate the ray distortions induced by stellar mass distributions, so they are not limited by any sort of resolution effects. The difference must therefore come down to differences between the stellar distributions found in \eagle and those used by \citetalias{Hilbert2008}. \citetalias{Hilbert2008} modelled the stellar component as the sum of an exponential disc
\begin{equation}
\Sigma_\mathrm{disc} = \frac{M_\mathrm{disc}}{2 \pi r_\mathrm{disc}^2} \exp \left( - \frac{r}{r_\mathrm{disc}} \right)
\end{equation}
and a bulge
\begin{equation}
\Sigma_\mathrm{bulge} = \frac{94.5 \, M_\mathrm{bulge}}{r_\mathrm{bulge}^2} \exp \left[ - 7.67 \left(\frac{r}{r_\mathrm{bulge}}\right)^{1/4} \right].
\end{equation}
The disc mass, bulge mass and disc radius ($M_\mathrm{disc}$, $M_\mathrm{bulge}$ and $r_\mathrm{disc}$ respectively) were taken from the 
\citet{2007MNRAS.375....2D} semi-analytic model, that had been run on merger trees generated from the same Millennium simulation that \citetalias{Hilbert2008} used for the DM component. They used an observationally derived $M_\mathrm{bulge} - r_\mathrm{bulge}$ relation to get the bulge radius.

To verify that these parametric mass distributions, with values taken from \citet{2007MNRAS.375....2D}, produce less strong lensing than \eagle galaxies, we use the Millennium database\footnote{\href{http://gavo.mpa-garching.mpg.de/Millennium/}{http://gavo.mpa-garching.mpg.de/Millennium/}, described in \citet{2006astro.ph..8019L}. Strictly speaking we use the milli-Millennium database, similar to the Millennium database but openly accessible, and for a simulation with a volume 1/512 of the full Millennium simulation.} to find the \citet{2007MNRAS.375....2D} galaxy parameters for haloes with $10^{12} < M_{200} / \msun < 10^{13}$ in the $z=0.362$ snapshot. For each halo, we calculate $\sigma_\mathrm{lens}^\mathrm{S}$ for $|\mu| > 10$, which we can then compare with the \eagle haloes in Fig.~\ref{fig:convergence}. We find that at fixed halo mass, the median $\sigma_\mathrm{lens}^\mathrm{S}$ from this procedure is one order of magnitude lower than from \eagle, which is true for both $10^{12}$ and $10^{13} \msun$ haloes. This comparison is slightly unfair, because we have not included the DM component in the Millennium lensing. At $10^{13} \msun$ (where \eagle and \citetalias{Hilbert2008} differ by a factor of two to three in $\partial \tau / \partial \log M$) the DM fraction within the Einstein radius is typically significant, whereas at $10^{12} \msun$ the lensing is dominated by the stars. As such, the increasing discrepancy in $\partial \tau / \partial \log M$ going from $10^{13} \msun$ to $10^{12} \msun$ is explained by the increasing importance of the stars for the lensing, and the stellar components used by \citetalias{Hilbert2008} being less efficient lenses than those found in \eagle.

Given that \eagle has a stellar mass -- stellar size relation that is a good match to observations \citep{2015MNRAS.446..521S}, and a similar comparison with the stellar mass -- halo mass relation suggests that \eagle has too few stars in haloes around $10^{12} \msun$, it seems unlikely that \eagle is significantly overestimating the lensing contribution from $10^{12} \msun$ haloes. A thorough analysis of the \citet{2007MNRAS.375....2D} semi-analytic galaxies and their lensing signal as implemented by \citetalias{Hilbert2008} is beyond the scope of this work, but here we mention possible reasons for the lensing signals being lower than in \eagle. The simplest possibility is that there is simply not enough mass in stars, or that the galaxies are too large (and therefore more diffuse, and so less efficient strong lenses). Another possibility is that departures from circular symmetry may be important. In particular, \citetalias{Hilbert2008} assume the stellar disc is always seen face on. Most of their galaxies in the $10^{12} - 10^{13} \msun$ halo mass range are disc-dominated, and viewing these discs face-on their surface densities rarely exceed $\Sigma_\mathrm{crit}$. An edge-on disc reaches much higher surface densities and is therefore a more powerful lens \citep{1998ApJ...503...48B, 1999MNRAS.303..423B}. The choice to place all discs face-on may therefore cause \citetalias{Hilbert2008} to underestimate the contribution to the strong lensing optical depth from lower mass haloes.

\subsection{Comparison with lensed submillimeter galaxies}

As mentioned in the introduction, one of the arguments for strong lensing being dominated by galaxy lenses, is that these are the primary lensing population when galaxies detected as being the brightest at submillimeter wavelengths are followed up to allow identification of a potential lens \citep[for example in][]{2013ApJ...762...59W}. Here we sketch out a qualitative argument for why this is expected, and does not contradict our finding that lenses with $M_{200} > 10^{13} \msun$ make a dominant contribution to the strong lensing optical depth for high magnifications of point sources.

The very brightest objects observed in the submillimeter are almost entirely gravitationally lensed, which can be understood from the steepness of the bright-end of the intrinsic luminosity function of dust-obscured star-forming galaxies \citep{2002MNRAS.329..445P, 2010ApJ...717L..31L}. This means that a small fraction of less-bright objects (of which there are many) being highly magnified, can dominate over the intrinsically bright objects with the same observed flux \citep{2010Sci...330..800N}.\footnote{This is the same argument being made by \citet{Broadhurst2018}, who, by assuming that the mass function of black holes exponentially decreases at masses $> 10 \msun$, find that objects detected as having masses $\sim 30 \msun$ would in fact be dominated by lensed objects that are intrinsically less massive.}

The flux limits employed in submillimeter surveys to find likely lensed objects, primarily select for objects with only modest magnifications. For example, \citet{2013ApJ...762...59W} expect magnifications of around 9 given their sample cuts. Given the universal form of $P(> \mu) \propto \mu^{-2}$ for large $\mu$, if a class of objects dominates lensing for large magnifications (greater than 100 say), then it should also dominate lensing for more modest magnification (such as those relevant for submillimeter galaxies). However, this relation is true only for point sources, with extended sources having more complicated magnification distributions.

It is perhaps intuitive that with increasing source size, the maximum magnification achievable decreases, as less of the source can lie close to a caustic. A less obvious fact is that this decrease in maximum magnification is accompanied by an increase in the probability of being moderately magnified. This can be understood from noting that for an extended source, the magnification of the source as a whole is a weighted mean of the source plane magnification for point sources over the surface brightness profile of the extended source. As such, the mean magnification of sources randomly distributed on the source plane must be independent of their size. A large source cannot all be close to a caustic, but there is an increased chance that at least some of it will be. This effect was calculated explicitly by \citet{2018MNRAS.481.2189D}, who showed that for a circular source with constant surface brightness lensed by an SIS, the cross-section for $|\mu| > 10$ is maximised when the angular radius of the source is 30\% of the Einstein radius. As galaxies have smaller Einstein radii than clusters, the relative size of a submillimeter source is larger when lensed by a galaxy than by a cluster. This decreases the importance of galaxy-lenses for very high magnifications of extended sources, but increases their prevalence as lenses with moderate magnifications, which is what dominates observed submillimeter samples.

\subsection{Implications for searches for lensed gravitational waves}

Aside from its importance for trying to estimate the probability of strong lensing from observed galaxies or galaxy clusters, knowing which halo masses are responsible for strong lensing is important for strategies to find optical counterparts to gravitationally lensed gravitational waves. At present, gravitational wave detections have large positional uncertainties, with 90\% confidence regions typically covering a few hundred square degrees \citep{2018LRR....21....3A}. Surveying this whole area with optical telescopes requires a large number of exposures to \emph{tile} the sky localisation region. 

Reducing the telescope time required to find optical counterparts to gravitational waves requires novel observing strategies. For example, the optical counterpart to the binary neutron star merger detected by LIGO and Virgo \citep[GW170817,][]{PhysRevLett.119.161101} was first discovered \citep{2017Sci...358.1556C} using an observing strategy that targeted known galaxies in the three-dimensional LIGO-Virgo localisation \citep{2016ApJ...820..136G}.

For high redshift gravitational wave sources, narrowing down the search using plausible host galaxies will be difficult, firstly because a larger fraction of possible hosts will be undetected and secondly because of the large number of galaxies per unit solid angle at high redshift compared with low redshift. However, if there is evidence that a particular source may have been lensed, then a search strategy that targets lenses rather than source-hosts can be used. Such a strategy has been employed by \citet{2019MNRAS.485.5180S}, who observed two known strong lensing clusters within the sky localisation of a binary black hole seen by LIGO-Virgo \citep[GW170814,][]{PhysRevLett.119.141101}.

Assuming that an observed gravitational wave has been strongly lensed, and that there is an electromagnetic counterpart to detect, for the \citet{2019MNRAS.485.5180S} strategy to have a high chance of success requires that massive clusters dominate the strong lensing optical depth. From $\partial \tau / \partial \log M$ in Fig.~\ref{fig:ODSL_combo}, we can see that the most massive clusters contribute only a small fraction of the total optical depth. In fact, using the \bahamas + \eagle prediction, if we take a $z=2$ point source that we know to have been highly magnified, the probability that it was lensed by a halo with $M_{200} > 10^{15} \msun$ is only around 2\%, rising to 25\% for $M_{200} > 10^{14} \msun$ and 50\% for $M_{200} > 10^{13} \msun$. However, a single powerful strong-lensing cluster can have a source-plane cross-section for $|\mu|>10$ of $\sim 500 \, \rm arcsec^2$.
\footnote{As a concrete example, MACS~0717 has $\sigma_\mathrm{lens}^\mathrm{S}(|\mu|>30) = 65 \, \rm arcsec^2$ \citep{2019MNRAS.486.5414V}. Converting this to $|\mu|>10$ using $P(|\mu|>\mu_0) \propto 1/\mu_0^2$ gives a cross-section of $580 \, \rm arcsec^2$.} 
This would account for around 5\% of the highly magnified source plane solid angle within the sky localisations of the best localized GW detections (90 per cent localisation uncertainty of $\sim50\rm \, degree^2$), if indeed such a cluster lens is located within the sky uncertainties.  In general, our results indicate that finding optical counterparts to high-redshift strongly lensed GWs will most likely require very wide-field optical follow-up as discussed by \citet{2019arXiv190205140S}.  Full exploration of such wide-field follow-up will benefit from highly complete lists of strong-lensing systems down to halo masses of $10^{13}M_\odot$ \citep{2020MNRAS.495.1666R}.

\subsection{Strong lensing probabilities at different source redshifts}

The mass and redshift distribution of gravitational lenses is important for designing strategies to find the likely lenses of a lensed gravitational wave, as we have discussed. However, for calculating the rate of detectable lensed events, or for finding the probability that a given event has been strongly lensed (assuming one does not have an accurate sky localisation), the important quantity is simply the probability of strong lensing as a function of source redshift. \citet{Ng2018} have made predictions for the rate at which lensed gravitational waves should be detected, while \citet{2019ApJ...874L...2H} recently found no evidence that any observed gravitational wave signals have been strongly lensed. Both groups used a strong lensing optical depth based on the assumption that lensing was done by a population of SIS mass profiles, whose mass function does not evolve, and has a normalisation determined from galaxy surveys \citep{1991MNRAS.253...99F}.

\begin{figure}
        \centering
        \includegraphics[width=\columnwidth]{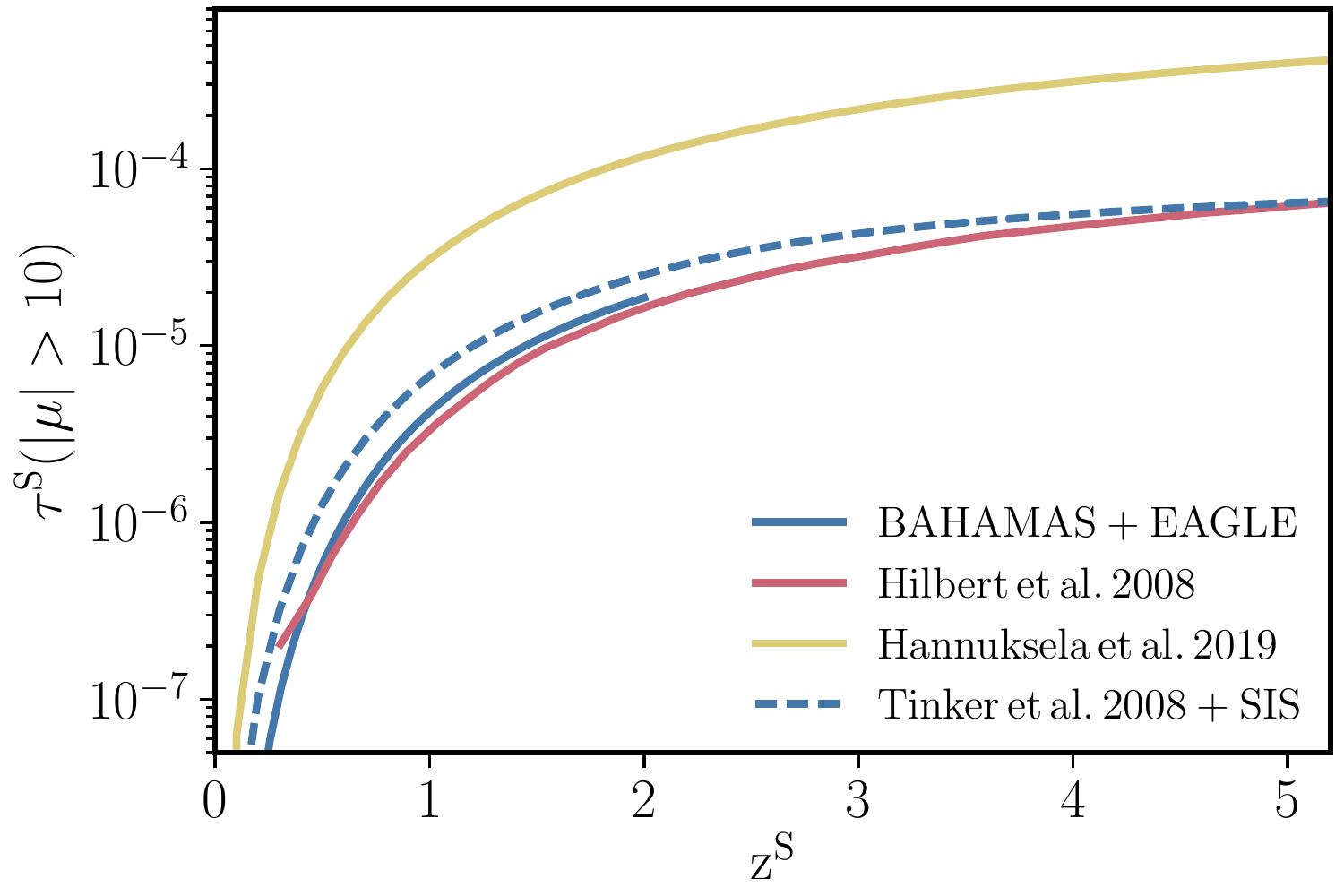}
	\caption{The source-plane optical depth for magnifications greater than 10, as predicted by different models. The solid blue line shows our prediction from hydrodynamical simulations, while the blue dashed line shows the prediction from our SIS model. The red line shows results from  \citetalias{Hilbert2008}, who also used cosmological simulations, but using a semi-analytic model of galaxy formation rather than hydrodynamical simulations. The yellow line shows the results from the SIS model used by \citet{2019ApJ...874L...2H}, in which the mass function of lenses does not change with redshift. The different predictions agree to within a factor of two over most of the redshift range, except for the \citet{2019ApJ...874L...2H} line which is a factor of four to six larger than the others.}
	\label{fig:tau_zs}
\end{figure}

In Fig.~\ref{fig:tau_zs} we plot the evolution of the source-plane optical depth for high magnification lensing as predicted from our simulations. This was done by calculating $\partial^2 \tau / \partial \log M \, \partial z$, as shown for $\zs = 2$ in Fig.~\ref{fig:ODSL_combo}, but at many different source redshifts. At each $\zs$, we integrate over lens mass and lens redshift to obtain $\tau(\zs)$. As we only generated mass maps for $\zl \leq 2$, we can only calculate the optical depth out to $\zs = 2$. For comparison we show this same quantity from using our SIS model and from \citetalias{Hilbert2008}. We find that the two simulation predictions agree well with one another, while the SIS model predicts slightly more strong lensing, particularly at low source redshifts. We also used the SIS model to understand the impact of a change in cosmology from that adopted in our simulations \citep{2014A&A...571A...1P} to the latest \emph{Planck} cosmology \citep{2018arXiv180706209P}. This latest cosmology has a slightly increased $\Omega_\mathrm{m}$ and a decreased $\sigma_8$ compared with the earlier \emph{Planck} cosmology. These changes approximately cancel one another to produce a $\tau(\zs)$ that differs from the \citet{2014A&A...571A...1P} SIS result in Fig.~\ref{fig:tau_zs} by less than the line width (the \citet{2018arXiv180706209P} line is not plotted in Fig.~\ref{fig:tau_zs}).

To compare with work that was used to calculate the probabilities that observed gravitational waves have been strongly lensed, we also plot the optical depth used by \citet{2019ApJ...874L...2H}. They used a model where the population of lenses was comprised of singular isothermal spheres, with a mass function that does not evolve with time. This leads to an optical depth as a function of source redshift
\begin{equation}
\tau (\zs, |\mu| \ge \mu_0) = F \left( \frac{d_\mathrm{C}(\zs)}{c H_0^{-1}} \right)^3 \left( \frac{2}{\mu_0} \right)^2,
\label{eq:tau_zs_general}
\end{equation}
where $d_\mathrm{C}(z)$ is the comoving distance to redshift $z$, and $F$ is a dimensionless constant for which they use 0.0017. This is plotted in Fig.~\ref{fig:tau_zs} for $\mu_0 = 10$, along with the same quantity predicted from our simulations, our SIS model and from \citetalias{Hilbert2008}.

In Fig.~\ref{fig:tau_zs}, the different models agree reasonably well on $\tau(\zs)$, except for that used by \citet{2019ApJ...874L...2H} which predicts more strong lensing than the others by a factor of approximately five. This means their value for $F$ is probably optimistic in terms of the amount of strong lensing, but a lower value for $F$ would only strengthen their conclusion that it is unlikely that any of the already detected gravitational wave signals have been highly magnified by gravitational lensing.

\section{Conclusions}
\label{sect:conc}

In this paper we have calculated the optical depth for high magnification gravitational lensing of high-redshift point sources, and its contribution from different lens masses and redshifts, as predicted by two recent hydrodynamical simulations, \eagle and \bahamas. We combine these two simulations because \eagle has sufficient resolution to resolve the lowest mass haloes important for strong lensing, while the large volume of \bahamas allows for an adequate sampling of the high-mass end of the halo mass function. The predicted contribution from group mass ($10^{13} - 10^{14} \msun$) haloes is rather uncertain, because the strong lensing regions of these haloes are not well resolved in \bahamas, and the \eagle simulation has insufficient volume for a robust prediction. Future simulations, with sufficiently large volume and resolution to accurately estimate the lensing contribution from galaxy groups, will improve our predictions further.

The relative contribution of different lens masses to the total optical depth for large magnifications has been a topic of recent debate, particularly in the community studying the effects of gravitational lensing on observations of gravitational waves. The importance of galaxies versus galaxy clusters is a `tug-of-war' between the high number density of galaxies, and the large lensing cross-sections of galaxy clusters. We first studied a simple model, in which haloes are modelled as singular isothermal spheres. In this model, the lensing cross-section of haloes grows as $M^{4/3}$, while the number density of haloes per decade in halo mass is approximately proportional to $M^{-1}$. As such, the high lensing cross-sections of more massive haloes wins over the increasing number of less massive haloes, and it is the more massive haloes that contribute more to the strong lensing cross-section. This argument holds up to the mass scale at which the mass function is no longer a power-law, but is exponentially suppressed, which reduces the importance of the most massive haloes for strong lensing.  

The hydrodynamical simulations confirm this picture, while altering slightly the importance of different halo masses. The primary difference between the simulation predictions and those from our SIS model is that the simulations pick out a particular mass scale ($M_{200} \sim 10^{12} \msun$) as being more efficient at lensing than predicted by the SIS model. This scale corresponds to the scale at which galaxy formation is most efficient, in that the stellar mass to halo mass ratio is the highest there \citep{2013MNRAS.428.3121M}. A second result of the hydrodynamical simulations is that they predict a minimum halo mass below which lensing becomes inefficient. Again this is related to the stellar to halo mass relationship, as in low mass haloes it is the stars rather than dark matter that dominate the strong lensing region. The stellar mass falls off quickly with decreasing halo mass below a halo mass of $10^{12} \msun$, and so lower mass haloes quickly become unimportant for lensing. Overall we find that around half of all high magnification lensing is done by haloes with $M_{200} > 10^{13} \msun$ (i.e. galaxy groups and clusters), with the other half coming from less massive systems (galaxies). This result differs somewhat from a previous simulation-based result \citepalias{Hilbert2008}, which used a dark matter only simulation combined with analytical gravitational potentials for the stellar components of galaxies, in that we find an enhanced contribution from galaxies living in $10^{12} - 10^{13} \msun$ haloes.


We also discussed the implications of this work for strategies to hunt for optical counterparts to gravitationally lensed gravitational waves. If the bulk of high magnification lines of sight resulted from the most massive haloes, then a credible strategy for finding the optical counterpart to a gravitational wave that was known to be strongly lensed, would be to look in the strong lensing region of the most massive haloes within the gravitational wave sky localisation. However, massive clusters (with $M_{200} > 10^{15} \msun$) contribute only $2\%$ of the total optical depth for large magnifications, with the bulk of the signal coming from massive galaxies or low-mass clusters. The number densities of these objects is much higher, such that there will be many within the sky localisation of a detected gravitational wave. This means that finding the optical counterparts to high-redshift gravitationally lensed gravitational waves will most likely require tiling the credible area on the sky as determined from the gravitational waves. That said, if a particularly powerful strong lensing cluster was in the sky localisation of a well-constrained gravitational wave ($\sim50\rm \, degree^2$) then it could account for around 5\% of the strong lensing within the sky localisation. 

Finally, we presented our prediction for the optical depth for high magnification as a function of source redshift, a key ingredient in calculating the expected rate of gravitationally lensed gravitational waves, as well as the abundance of other gravitationally lensed point sources. We found that our result was in reasonable agreement both with \citetalias{Hilbert2008}'s previous simulation-based result, as well as simple models that treat lenses as SISs. At low redshift, high magnifications are extremely unlikely. Each $\zs=0.5$ source produces, on average, $\num{5e-7}$ images with magnification greater than ten; this increases to about $\num{2e-5}$ images for $\zs =2$. These lensing probabilities are lower than assumed by  \citet{2019ApJ...874L...2H} in recent work on the probability that observed gravitational waves have been strongly lensed. They found that strong lensing was unlikely to have affected the current sample of observed gravitational waves, and our lower intrinsic lensing probabilities strengthen this result.

\section*{Acknowledgments}

We give special thanks to Stefan Hilbert for producing new lensing results from the Millennium simulation, and to Ian McCarthy for granting us access to the particle data from the \bahamas simulations. Thanks also to Christopher Berry for helping with the early stages of this work and for providing comments on a draft version of this paper, and to James Nightingale, David Lagattuta, Joop Schaye, Otto Hannuksela and Thomas Callingham for helpful discussions.

AR is supported by the European Research Council (ERC-StG-716532-PUNCA) and the STFC (ST/N001494/1). GPS and MB acknowledge support from STFC grant number ST/S000305/1. RM acknowledges the support of a Royal Society University Research Fellowship. VRE acknowledges support from STFC grant ST/P000541/1. MJ is supported by the United Kingdom Research and Innovation (UKRI) Future Leaders Fellowship `Using Cosmic Beasts to uncover the Nature of Dark Matter' (MR/S017216/1). 

This work used the DiRAC@Durham facility managed by the Institute for
Computational Cosmology on behalf of the STFC DiRAC HPC Facility
(www.dirac.ac.uk). The equipment was funded by BEIS capital funding
via STFC capital grants ST/K00042X/1, ST/P002293/1, ST/R002371/1 and
ST/S002502/1, Durham University and STFC operations grant
ST/R000832/1. DiRAC is part of the National e-Infrastructure.

\bibliographystyle{mnras}
\bibliography{bibliography}

\bsp
\label{lastpage}

\end{document}